\documentclass[fleqn,usenatbib]{mnras}

\usepackage{newtxtext,newtxmath}

\usepackage[T1]{fontenc}
\usepackage{ae,aecompl}

\usepackage{graphicx}	
\usepackage{amsmath}	
\usepackage{amssymb}	

\title[Open cluster proper motions]{Update of membership and mean proper motion of open clusters from UCAC5 catalog}

\author[W. S. Dias et al.]{
W. S. Dias,$^{1}$\thanks{E-mail: wiltonsdias@yahoo.com.br}
H. Monteiro$^{1}$ and 
M. Assafin,$^{2}$
\\
$^{1}$UNIFEI, Instituto de F\'isica e Qu\'imica, Universidade Federal de Itajub\'a, Av. BPS 1303 Pinheirinho, 37500-903 Itajub\'a, MG, Brazil\\
$^{2}$Universidade Federal do Rio de Janeiro, Observat\'orio do Valongo, Ladeira Pedro Antonio 43, 20080-090 Rio de Janeiro, RJ, Brazil
}

\date{Accepted 2018 May 25. Received 2018 May 7; in original form 2017 December 18}

\pubyear{2018}

\begin{document}
\label{firstpage}
\pagerange{\pageref{firstpage}--\pageref{lastpage}}
\maketitle

\begin{abstract}
We present mean proper motions and membership
 probabilities of individual stars for optically visible open clusters, which have been determined using data from the UCAC5 catalog. 
 This follows our previous studies with the UCAC2 and UCAC4 catalogs, but now using improved proper motions in the GAIA reference frame.
In the present study results were obtained for a sample of 1108 open clusters. 
  For five clusters, this is the first determination of mean proper motion, and for the whole sample, we present results with a much larger number of identified
  astrometric member stars than on previous studies. 
  It is the last update of our Open cluster Catalog based on proper motion data only. Future updates will count on astrometric, photometric and spectroscopic GAIA data as input for analyses.  

  \end{abstract}




\section{Introduction}

Accurate membership determination is crucial for astrophysical studies of open clusters. 
It depends on the quality of the data, the mathematical models used and the possibility to distinguish two populations in the observed field.

In the last few years we witnessed many advances in the membership determination both using kinematic and photometric data (\cite{Sampedro2017}, \cite{Monteiro2017}, \cite{Krone-Martins2014}, \cite{Dias2014}, among others). By other hand, astrometric data with unprecedented accuracy became available with the publication of the first version of data from the GAIA satellite \citep{Brown2016}. This allowed the analysis of nineteen nearby open clusters, as presented in \citet{vanLeeuwen2017}.

The GAIA DR2 catalog \citep{GAIA-DR22018} published parallax and proper motion data accurate enough to precisely distinguish the member stars of a cluster. Spectroscopic radial velocities were also published for bright nearby stars.
For this reason this is our last membership study using proper motion data only. In the forthcoming papers we will present memberships using kinematic, photometric and spectroscopic GAIA data.

\citet{Zacharias2017} published the UCAC5 catalog containing improved proper motions in the GAIA reference frame. The precision and the systematic errors in proper motions were improved by a factor of 2 with respect to the UCAC4 catalog \citep{Zacharias2013}. This new release motivated us to update the mean proper motions and membership probabilities of individual stars for the optically visible open clusters published before in \cite{Dias2014}.

This paper is organized as follows: in the next Section, we describe the data used. In Sect. 3, we briefly comment about the method and the procedures adopted for determining the membership probabilities and mean proper motion of the clusters. The results are shown in Sect. 4, while in Sect. 5 we present the comparison with the literature. We discuss and summarize the main results in the last section.

\section{Data used}

The UCAC5 catalog presents data in the GAIA reference frame for over 107 million objects covering the entire sky complete from the brightest stars to those with a magnitude about R = 16.
The formal errors in the proper motions of the stars range from about 1-2 mas yr$^{-1}$ for R = 11 - 15 mag, to about 5 mas yr$^{-1}$ at R=16. The  systematic errors in proper motions are estimated to be less than 1 mas yr$^{-1}$ for the entire magnitude range of the catalog.

We used the central coordinates and diameters of the open clusters published in our New catalog of Optically Visible Open Clusters and Candidates \citet{Dias2002}\footnote{The latest version (3.5) can be accessed on line at \url{https://wilton.unifei.edu.br/ocdb.}} to extract the data from the UCAC5 catalog\footnote{The extraction of the UCAC5 data was performed using the VizieR tool \url{http://vizier.u-strasbg.fr/viz-bin/VizieR?-source=I{\%}2F340}}. We complemented the input information using new central coordinates of ten open clusters updated by \citet{Sampedro2017}.  As done in the previous work, we opted to use a region in the sky
about 2 arcmin bigger than the area covered by the cluster to include virtually all possible members of the studied clusters. In the Table \ref{tab:results} we give the radius used for each cluster to extract the UCAC5 data.

\section{Method}

In this work we used exactly the same method presented in \citet{Dias2014} (hereafter D14).  
Briefly, the method consider the existence of two elliptical bivariate populations in the region (cluster and field stars) following \citet{Zhao1990} and the proper motion's errors in the frequency function as presented in Eq.\ref{eq:cluster} and Eq.\ref{eq:field}, where
the notation \textit{c} and \textit{f} subscripts for cluster and field parameters, respectively, x for the coordinate
$\mu_{\alpha} \cos \delta$, and y for the coordinate
$\mu_{\delta}$.
$\Phi = \Phi_c + \Phi_f$ is the total probability distribution,
$(\mu_{x,c},\mu_{y,c})$ are the averages of the cluster distribution with
standard deviations $\sigma_{x,c}$ and $\sigma_{y,c}$,
$(\mu_{x,f},\mu_{y,f})$ are the averages of the field distribution with
standard deviations $\sigma_{x,f}$ and $\sigma_{y,f}$, and $\rho_{c}$
and $\rho_{f}$ are the correlation coefficients of cluster and field
stars. The values $(\mu_{x},\mu_{y})$ are the component of the stellar proper motion, and
$\epsilon_{i}$ is the formal error in proper motion given by the
catalog.

\begin{eqnarray}
\label{eq:cluster}
\Phi_c(\mu_{x},\mu_{y}) =
  \frac{n_{c}}{2\pi\sqrt{\sigma_{x,c}^2+\epsilon_{i}^2}
    \sqrt{\sigma_{y,c}^2+\epsilon_{i}^2}
    \sqrt{1-\rho_{c}^2}} . . \\
. . X \exp\{ -\frac{1}{2(1-\rho_{c}^2)} [ 
\frac{(\mu_x - \mu_{x,c})^2}{\sigma_{x,c}^2+\epsilon_{i}^2}  . . \nonumber
\\ \  + . .
\frac{(\mu_y - \mu_{y,c})^2}{\sigma_{y,c}^2+\epsilon_{i}^2}
-2\rho_{c}(\frac{\mu_x-\mu_{x,c}}{\sqrt{\sigma_{x,c}^2+\epsilon_{i}^2}})
(\frac{\mu_y-\mu_{y,c}}{\sqrt{\sigma_{y,c}^2+\epsilon_{i}^2}})
] \}\ \ ,
\nonumber
\end{eqnarray}

\begin{eqnarray}
\label{eq:field}
\Phi_f(\mu_x,\mu_y) =
  \frac{1-n_{c}}{2\pi\sqrt{\sigma_{x,f}^2+\epsilon_{i}^2}
    \sqrt{\sigma_{y,f}^2+\epsilon_{i}^2}
    \sqrt{1-\rho_{f}^2}} . . \\
. . X \exp\{ -\frac{1}{2(1-\rho_{f}^2)} [ 
\frac{(\mu_x - \mu_{x,f})^2}{\sigma_{x,f}^2+\epsilon_{i}^2}  . . \nonumber
\\ \  + . .
\frac{(\mu_y - \mu_{y,f})^2}{\sigma_{y,f}^2+\epsilon_{i}^2}
-2\rho_{f}(\frac{\mu_x-\mu_{x,f}}{\sqrt{\sigma_{x,f}^2+\epsilon_{i}^2}})
(\frac{\mu_y-\mu_{y,f}}{\sqrt{\sigma_{y,f}^2+\epsilon_{i}^2}})
] \}\ \ 
\nonumber
\end{eqnarray}

  The probability density function for the whole sample is simply given by Eq.\ref{eq:probability}, where
{$n_c$ and $n_f$ are the number of cluster and field stars (non-members), respectively, normalized with respect to the total number of stars in the field.

\begin{eqnarray}
\Phi(\mu_x,\mu_y) = n_c \Phi_c(\mu_x,\mu_y) +
n_f \Phi_f(\mu_x,\mu_y)\ \ 
\label{eq:probability}
\end{eqnarray}

Note that $\rho$ is zero when the variables are stochastically independent. In the past it was common to transform the above function into a different coordinate frame in which the variables are stochastically independent. Basically this allowed to simplify the equation and facilitate computational work, which is no longer necessary (more details in \citet{Slovak1977} and \citet{Vasilevskis1965}).

In this work to obtain the unknown parameters (means, standard deviations, correlation coefficients, and numbers of members and non-members) we used the cross-entropy global optimization procedure by applying the maximum likelihood principle to the data considering their individual formal errors, which is listed in the proper motion catalog.
With the frequency function parameters we could determine
the individual probability of the membership of each star in the cluster by $P_{i} = \Phi_{c_{i}}/\Phi_{i}$.
We refer the reader the section 3.1 of D14 for a complete description since we used exactly the same procedure.

\section{Results}
The 2157 cataloged open clusters in the DAML02 with a diameter smaller than 300 arcmin were investigated and satisfactory results were obtained for a sample of 1108 clusters. 

In this work we opted to be very restrictive considering only the results which satisfies the following criteria:
\begin{itemize}
\item the number of stars in the field greater than twenty
\item the quality of the solution compatible with
the presence of two populations in the field, verified after visual inspection;
\item the final solution that provided parameters with numerical errors smaller than $1.0~ mas~yr^{-1}$.\end{itemize}

Table \ref{tab:results} (available in electronic form at CDS and in the DAML02 website) presents our final
results for the 1108 clusters and fields parameters.
Tables 2 to 1110, only available in electronic form, list the stars in the limits of each cluster, with the membership probabilities calculated by our method.

In this work we believe that the magnitude-dependent UCAC5 proper motion errors were already properly taken into account.

Typically, the estimated membership probability of each star decreases with the magnitude and it is small for mags $>$ 14-15. This is shown in Figure \ref{fig:membershipxmags} below. We noticed that the same behavior was observed for other fields in a very similar way to that presented in the Figure 3 of the paper published by \citet{Platais2011}.
This is an expected consequence of the method, since the individual stellar proper motion errors are considered in the estimated memberships as described in the Eq.\ref{eq:cluster} and in general these errors increase with the magnitude.

\begin{figure}
\centering
\includegraphics[scale = 0.6]{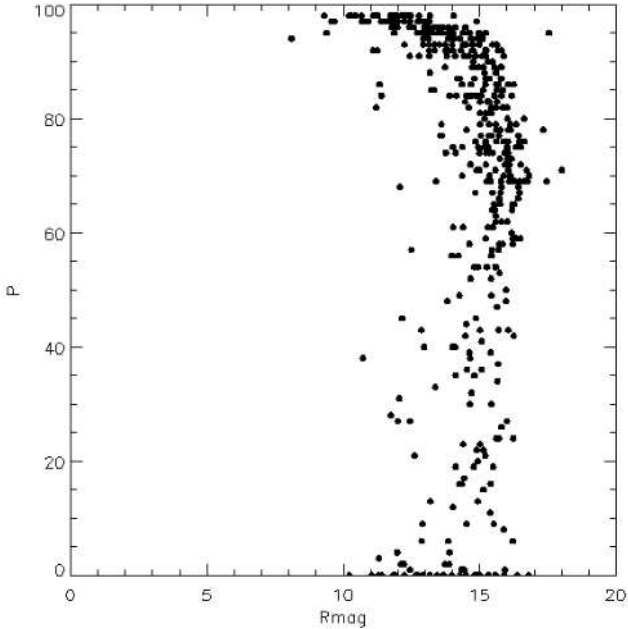}
\caption{The plot shows the estimated membership P given in \% as a function of Rmag for the stars in the field of cluster IC 4651.}
  \label{fig:membershipxmags}
\end{figure}

\begin{table*}
\caption[]{Example of the results of our proper motion analysis for open clusters. 
The meaning of the
symbols are as follows: 
R is the radius used to extract the data;
$N_{t}$ is the number of stars in the field;
$N_{c}$ is the number of cluster stars;
$N_{f}$ is the number of field stars;
$\mu_{\alpha}cos{\delta}$ and $\mu_{\delta}$ are the  proper
motion components in $mas~yr^{-1}$;
$\sigma$ is the dispersion of the proper motions components;
$\rho$ is the orientation angle of the minor axis of the elliptical cluster and field stars
proper motion distribution.}\label{tab:kin}
\begin{center}
\begin{tabular}{lcccccccc|ccccccc}
\hline
&\multicolumn{5}{c|}{Cluster} & \multicolumn{6}{c}{Field}\\
cluster       &   R     &    $N_{t}$  &  $N_{c}$      &    $\mu_{\alpha}cos{\delta}$     &   $\sigma$     &     $\mu_{\delta}$   &      $\sigma$    &     $\rho$    &     $N_{f}$      &  $\mu_{\alpha}cos{\delta}$    &    $\sigma$    &    $\mu_{\delta}$   &     $\sigma$   &      $\rho$   \\
\hline
Ruprecht 76   &  3.5   &    154  &     144    &       -4.68    &     1.25    &       2.68    &       1.37     &     0.03     &        10   &  -13.74   &      3.93   &      9.32   &       2.11  &        -0.25  \\
NGC 6178      &  3.5   &     289    &  241    &  -1.70   &     2.04     &     -3.11    &     1.64   &     0.01    &     48  &  -5.40  &       8.05    &      -6.37   &     10.60    &    -0.04   \\
\hline
\end{tabular}
\end{center}
\label{tab:results}
\end{table*}

\section{Comparison with the literature}

We compared the mean proper motions with those published in the literature to check the quality of the results. We used only large lists published after Hipparcos mission that cover about the same magnitude range. For the other cases we used the results
compiled in DAML02 catalog which is widely used. Although it is a heterogeneous list, we believe that it suffices for our purposes here, since for many clusters the selection of members was based on different criteria and data.

The quantity (D) used to compare the mean proper motions weighted by the formal errors 
is presented in the Eq.\ref{eq:D}, where CE refers
to the quantity obtained in this study from the global minimization method and LIT refers 
to the results obtained from literature. The mean value is shown in Tab.\ref{tab:compares}. 
The results obtained show almost all values lying below 2.5 to
3.0 indicating there is no statistical distinction between the distributions of proper motion 
that are compared within the estimated errors. 

\begin{eqnarray}
\label{eq:D}
D &=& \frac{(\mu_{\alpha \cos \delta_{CE}} - \mu_{\alpha \cos \delta_{LIT}})^{2}}{{ (\sigma_{\mu_{\alpha \cos \delta_{CE}}})^{2} + (\sigma_{\mu_{\alpha \cos \delta_{LIT}}})^{2} }} + \label{eqd} \\ \nonumber
&+&\frac{(\mu_{\delta_{CE}} - \mu_{\delta_{LIT}})^{2}} {{ (\sigma_{\mu_{\delta_{CE}}})^{2} + (\sigma_{\mu_{\delta_{LIT}}})^{2}}},
\nonumber  
\end{eqnarray}

Table \ref{tab:compares} and Figs. \ref{fig:compares1} and \ref{fig:compares2} give the results of the comparison with the literature in $\mu_{\alpha} \cos \delta$ and $\mu_{\delta}$. 
Essentially, the proper motions in the literature agree with the ones found in this work. There is 
no statistical distinction between the distributions 
since the differences found are small. This indicates that if systematic errors are present in our results, they are also present in same amounts in the values found in the literature. 

\begin{table*}
  \caption[]{Comparison of our results with those from the literature, which contains a large number of open clusters,  
   similar range in magnitude and were published after the Hipparcos mission. The differences are the values of this work minus literature presented in $mas~yr^{-1}$. The quantities D defined in the equation \ref{eq:D} are presented in the sixth column. This value gives the average of the  differences in the $\mu_{\alpha} \cos \delta$ and the $\mu_{\delta}$ components for each cluster, which are weighted by the formal errors. The last two columns give the number of common clusters and the catalog being compared, respectively. The references presented in the first column are discussed in the Section 5 of the text.}
\label{tab:compares}
\begin{center}
\footnotesize
\begin{tabular}{lccccccc}
\hline 
References & $\Delta{\mu_{\alpha \cos \delta}}$ & $\sigma{\Delta\mu_{\alpha \cos \delta }}$ &   $\Delta{\mu_{\delta}}$  & $\sigma{\Delta\mu_{\delta}}$ & D  &  N  &  catalog\\
\hline
D14    &  -0.6   &   2.0   &  0.2   &    2.2    &   1.4   &      1061    &   UCAC4\\
M1     &  -0.6   &   2.4   &  0.2   &    2.5    &   0.8   &      1032    &   UCAC4\\
M2     &  -0.7   &   2.2   &  0.2   &    2.4    &   0.3   &      1012    &   UCAC4\\
M3     &  -0.4   &   2.7   &  0.4   &    2.6    &   1.3   &      1064    &   UCAC4\\
K13    &   0.2   &   2.5   & -0.3   &    2.5    &   4.5   &      1032    &   PPMXL\\
DAML02 &  -0.4   &   3.0   &  0.1   &    2.4    &   4.0   &      1102    &    - \\
\hline
\end{tabular}
\end{center}
\end{table*}

\begin{figure*}
\centering
\includegraphics[scale = 0.38]{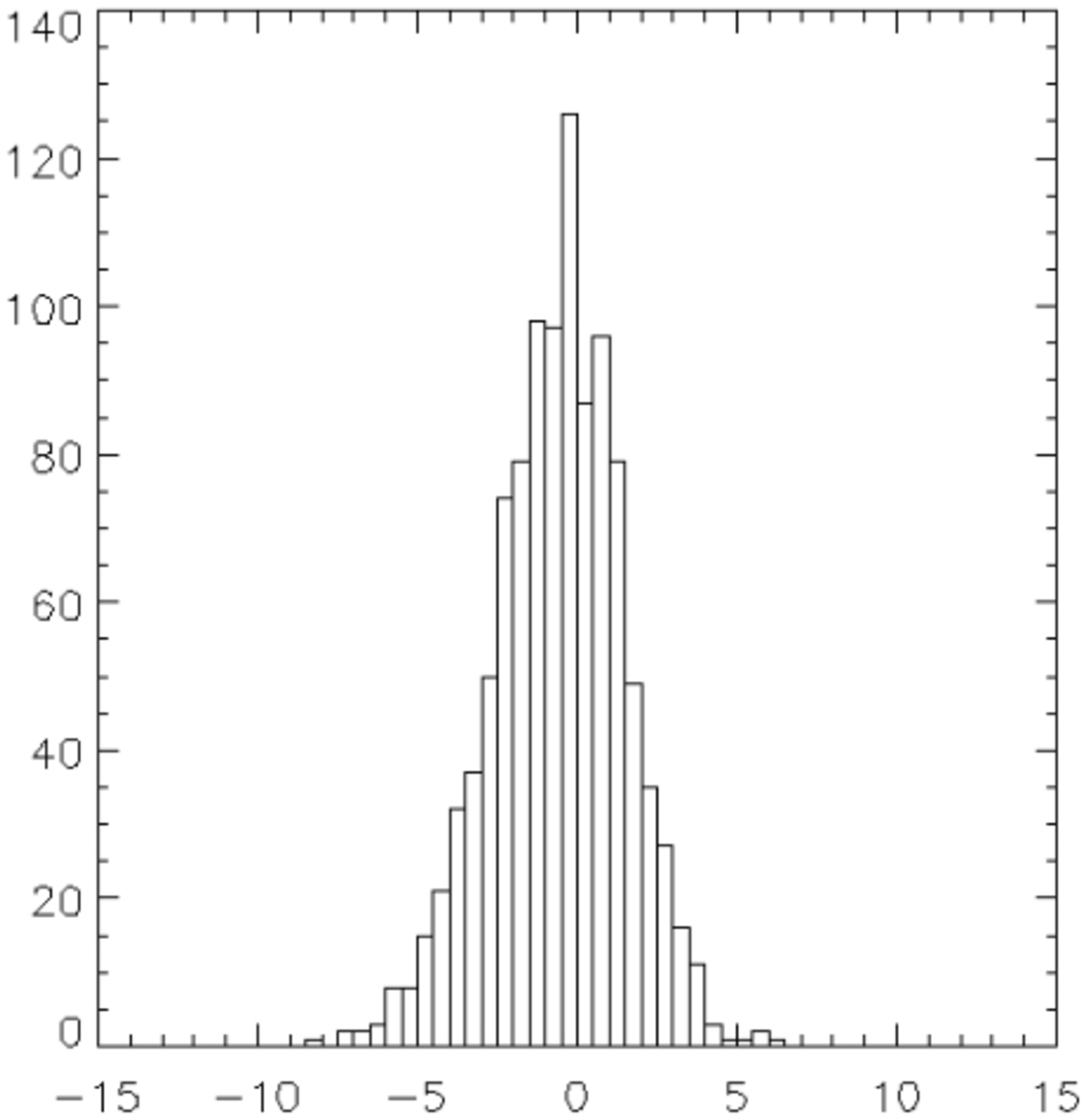}
\includegraphics[scale = 0.38]{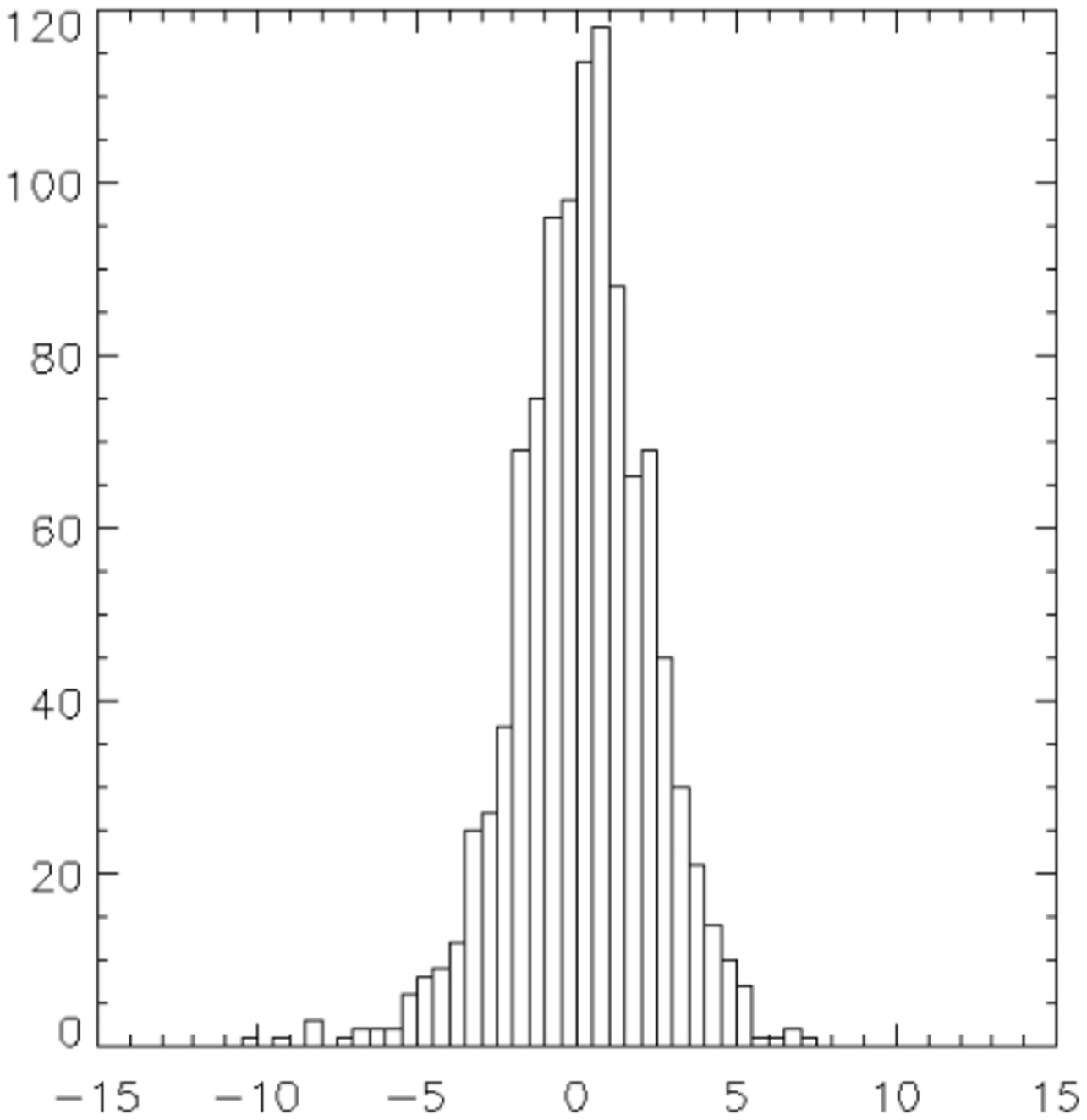}
\includegraphics[scale = 0.38]{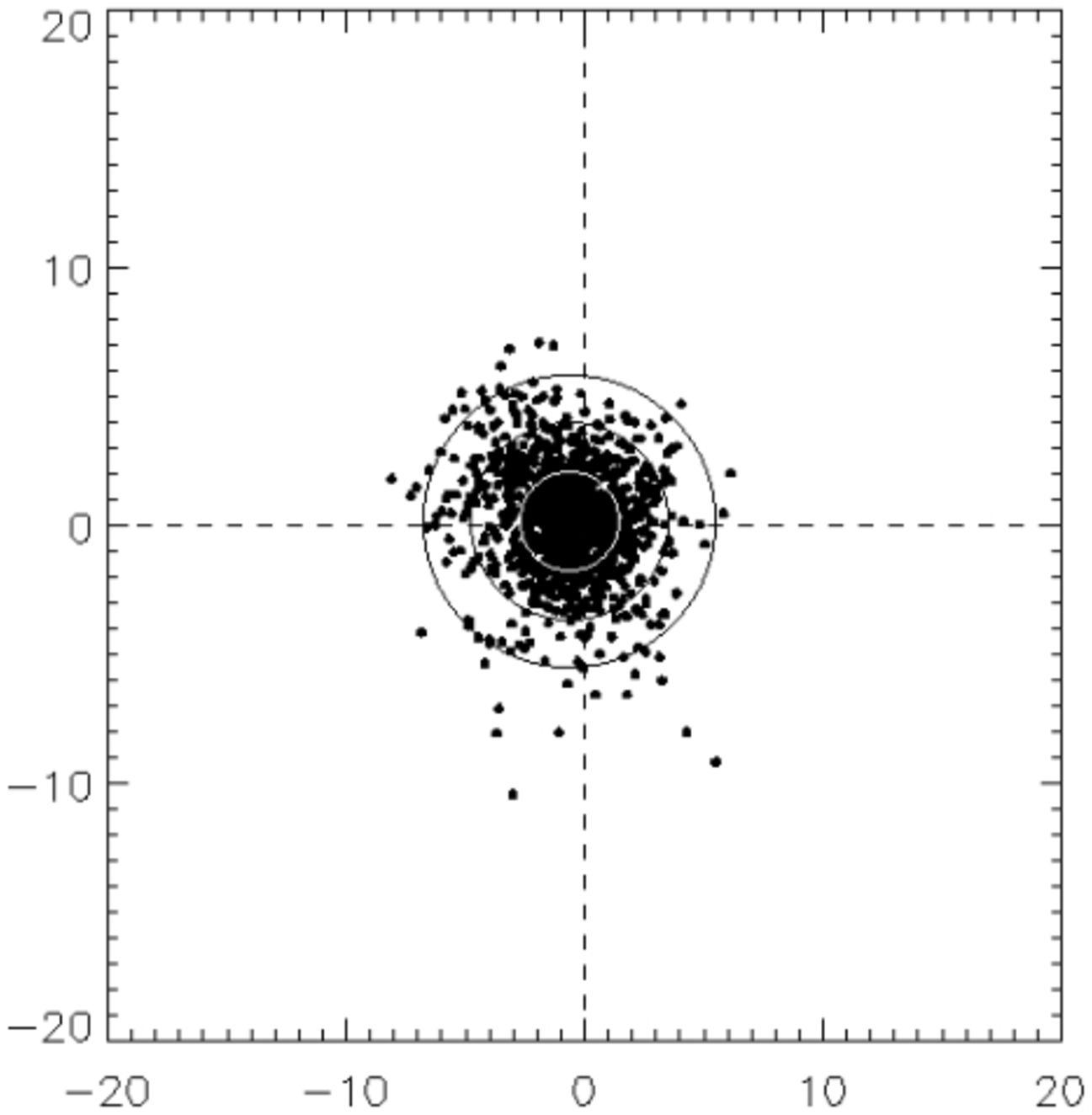}\\

\includegraphics[scale = 0.38]{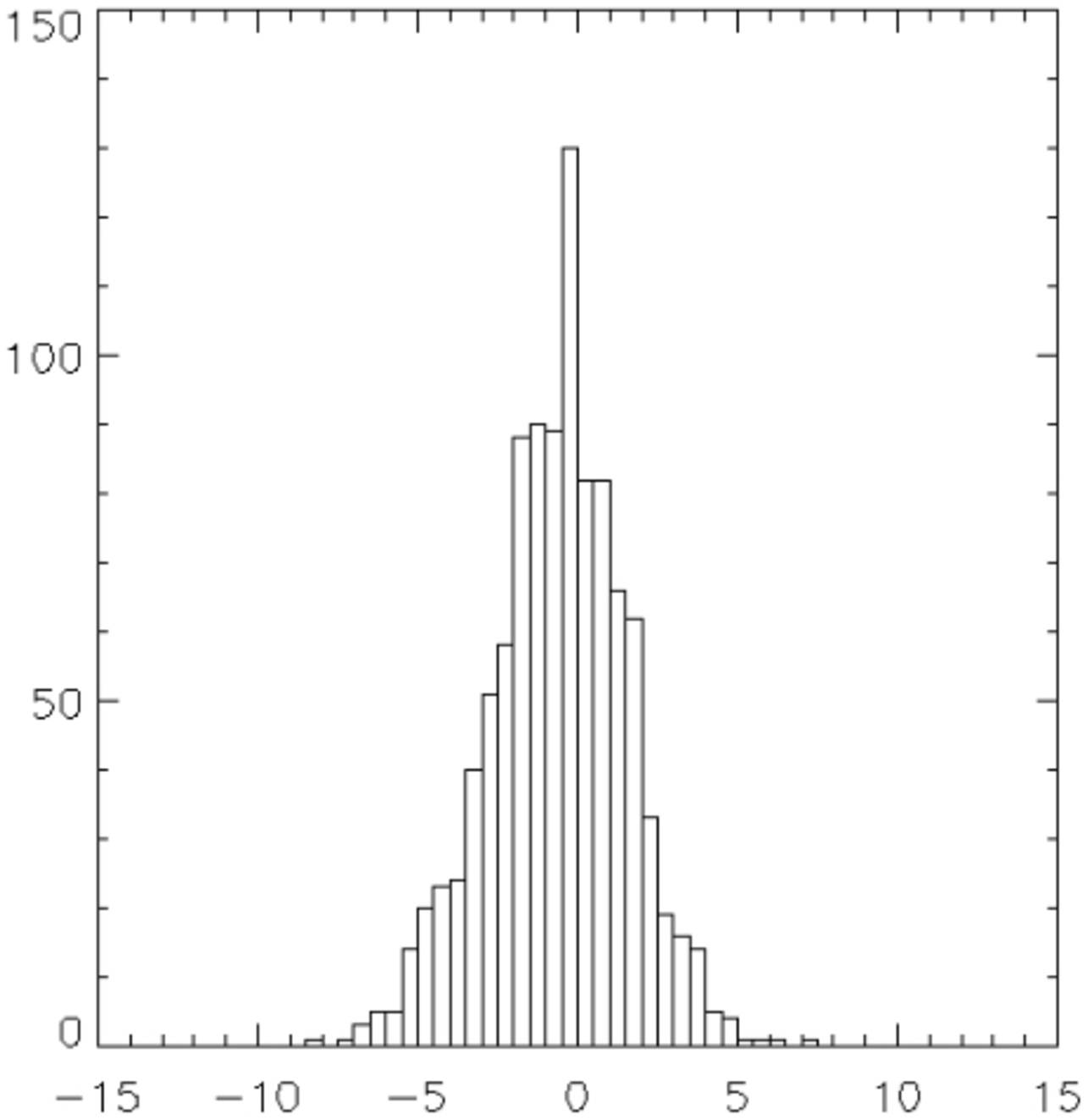}
\includegraphics[scale = 0.38]{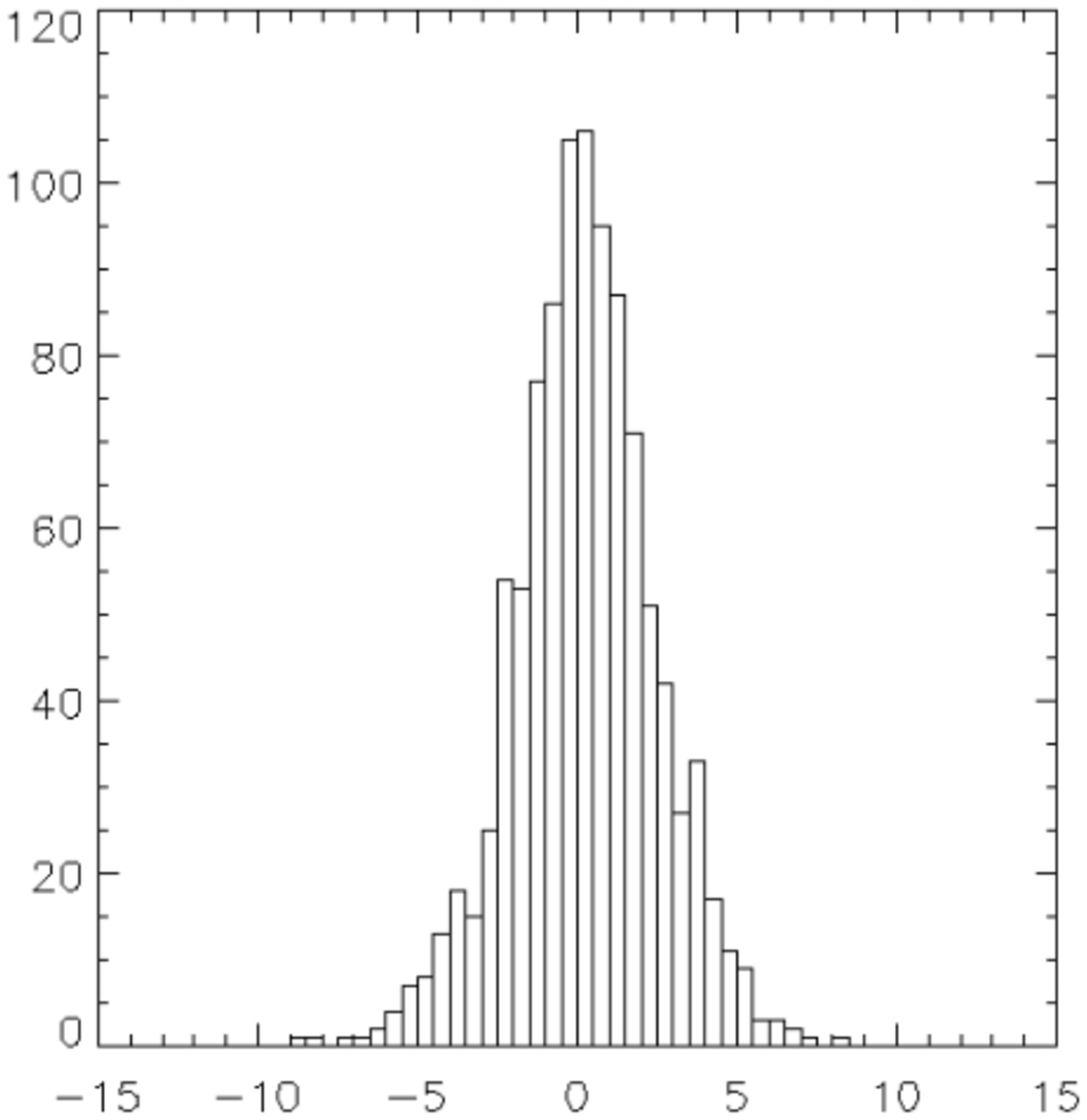}
\includegraphics[scale = 0.38]{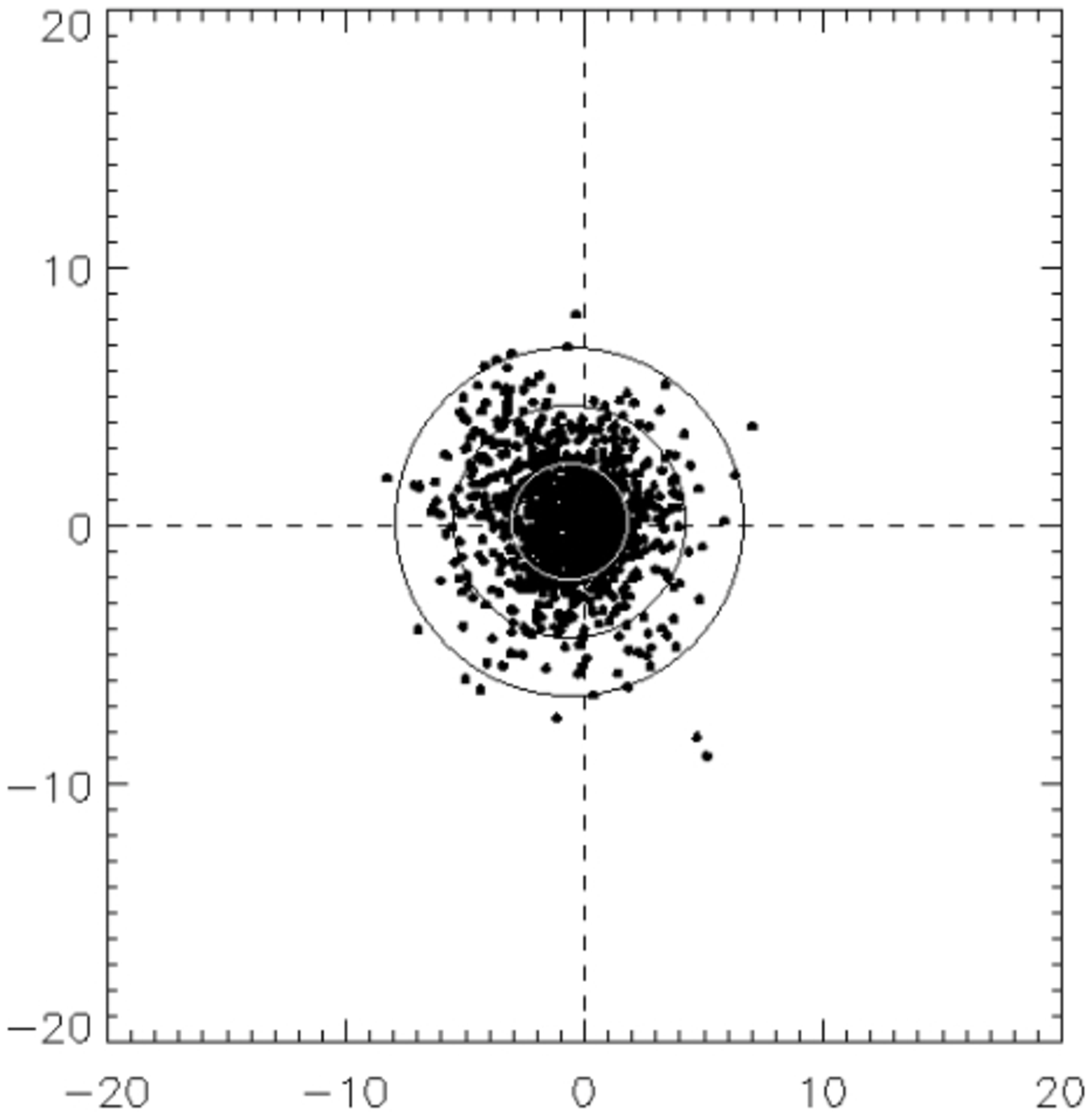} \\

\includegraphics[scale = 0.38]{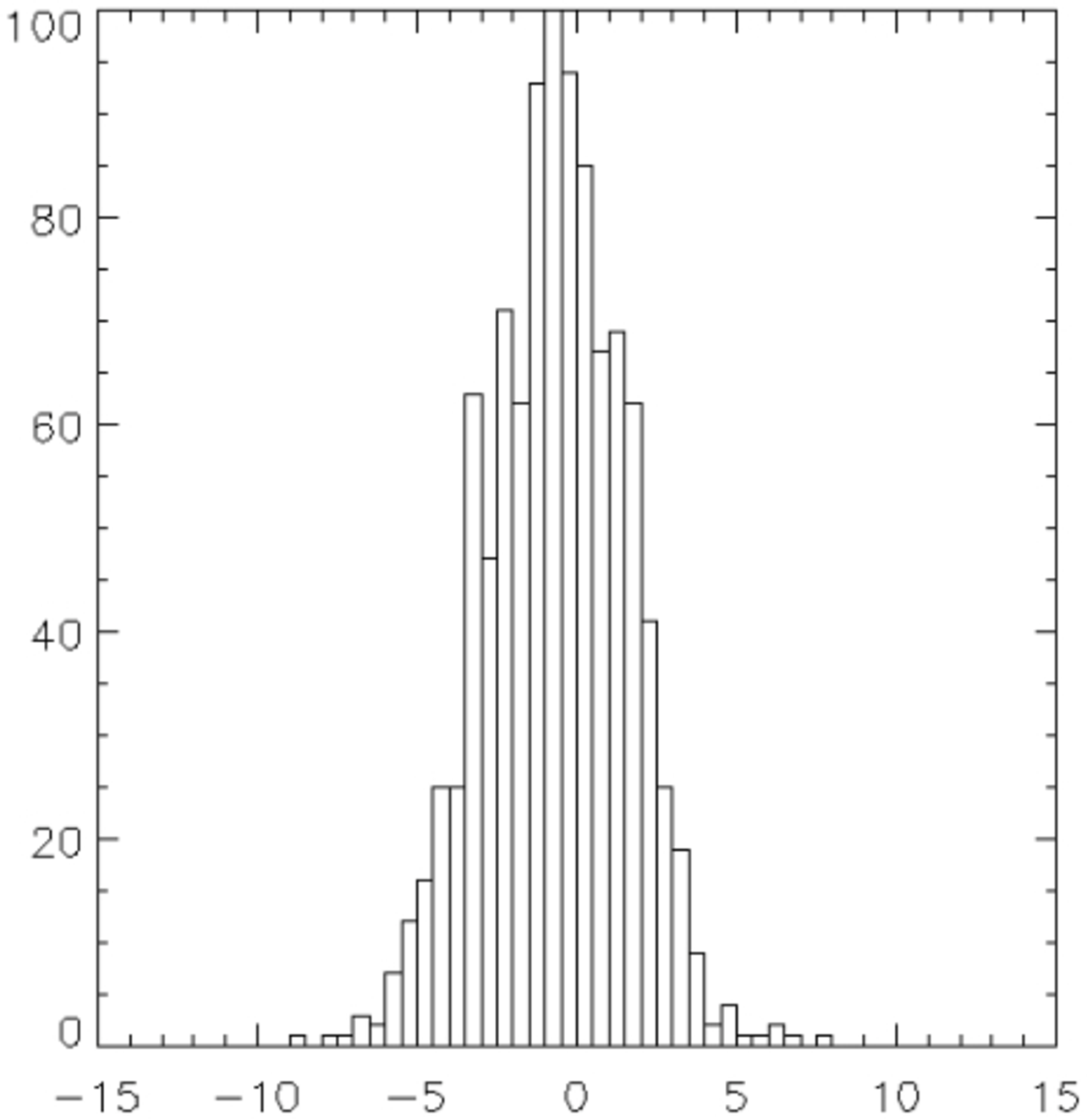}
\includegraphics[scale = 0.38]{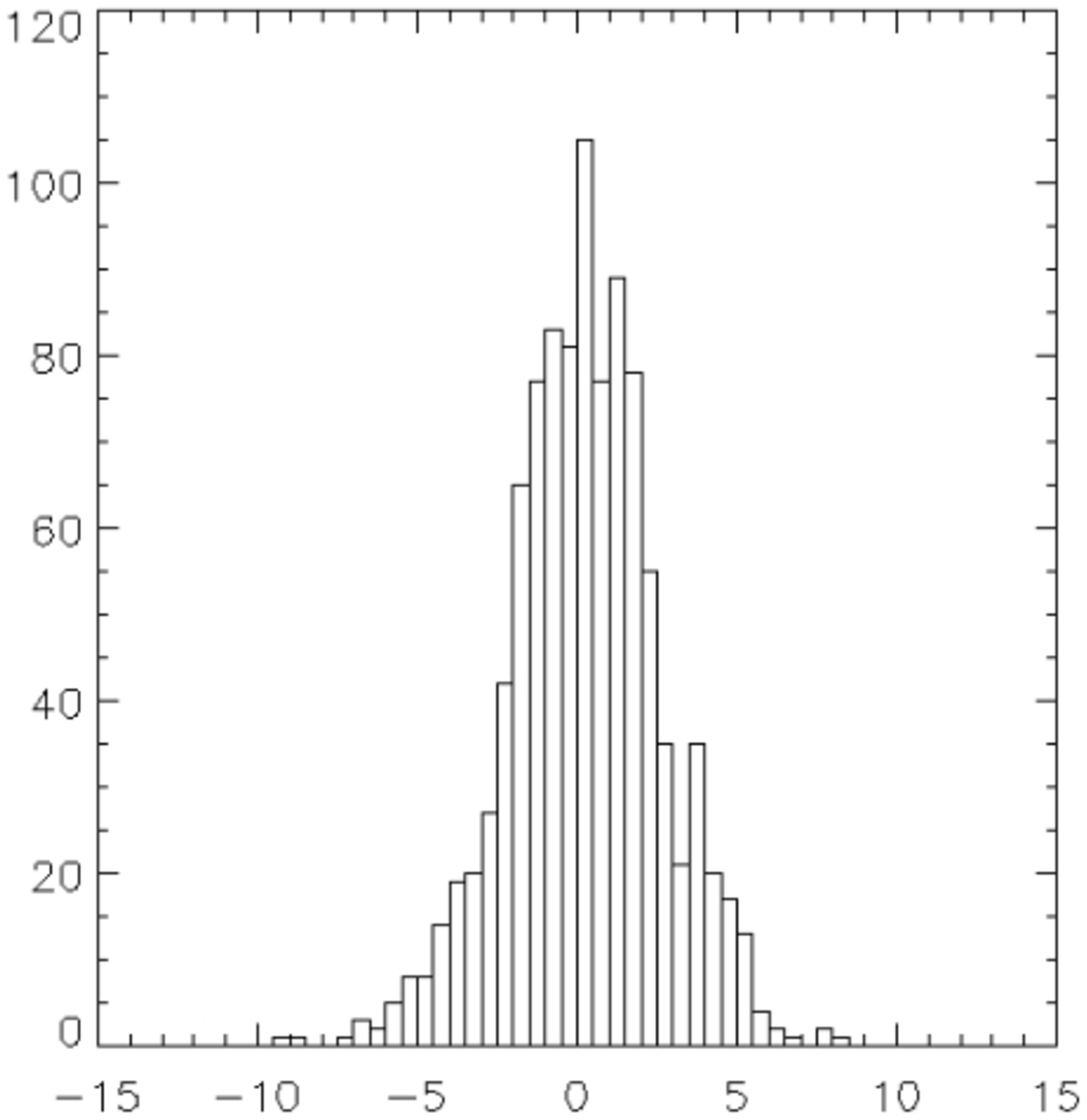}
\includegraphics[scale = 0.38]{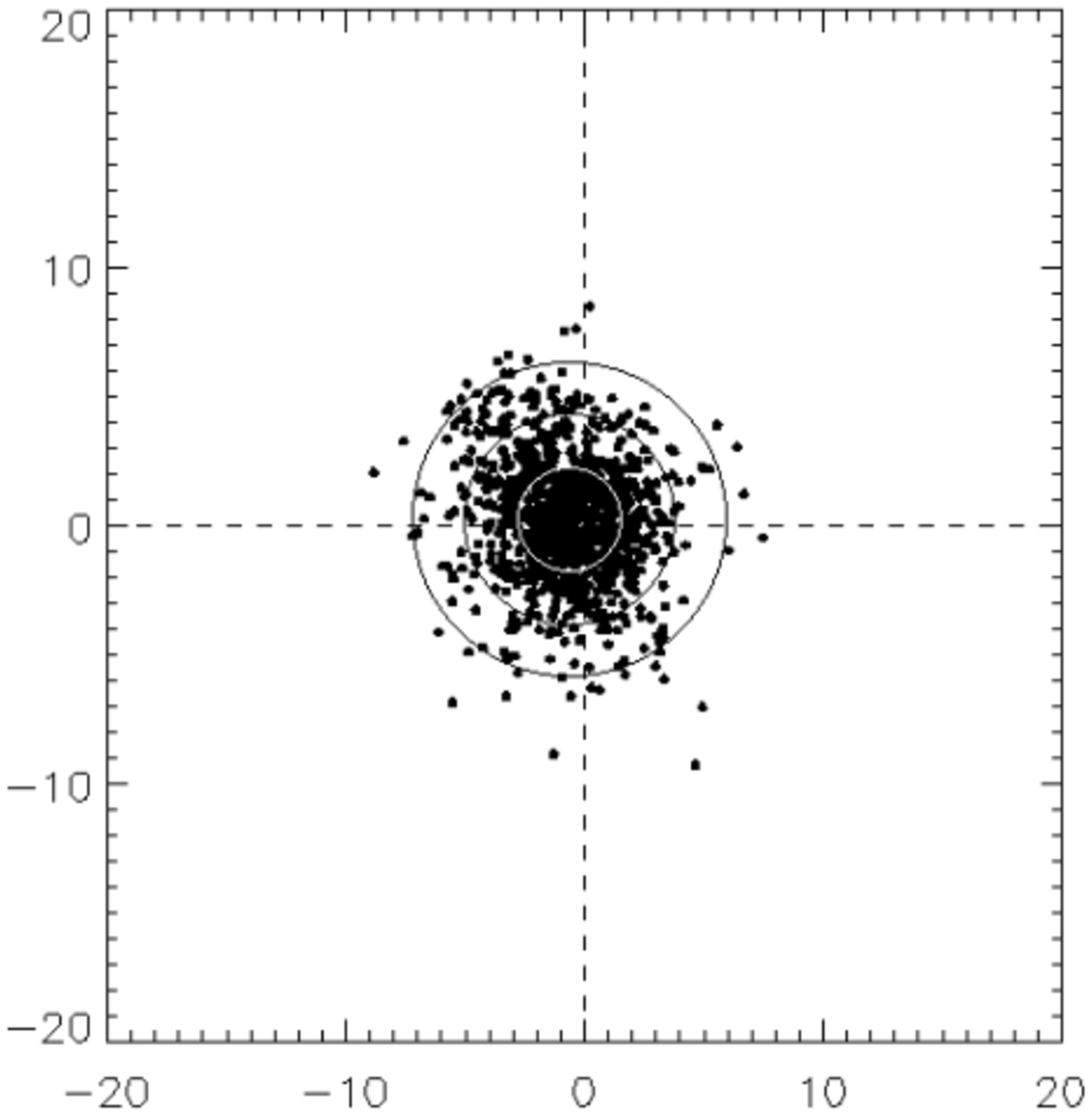}\\

\includegraphics[scale = 0.38]{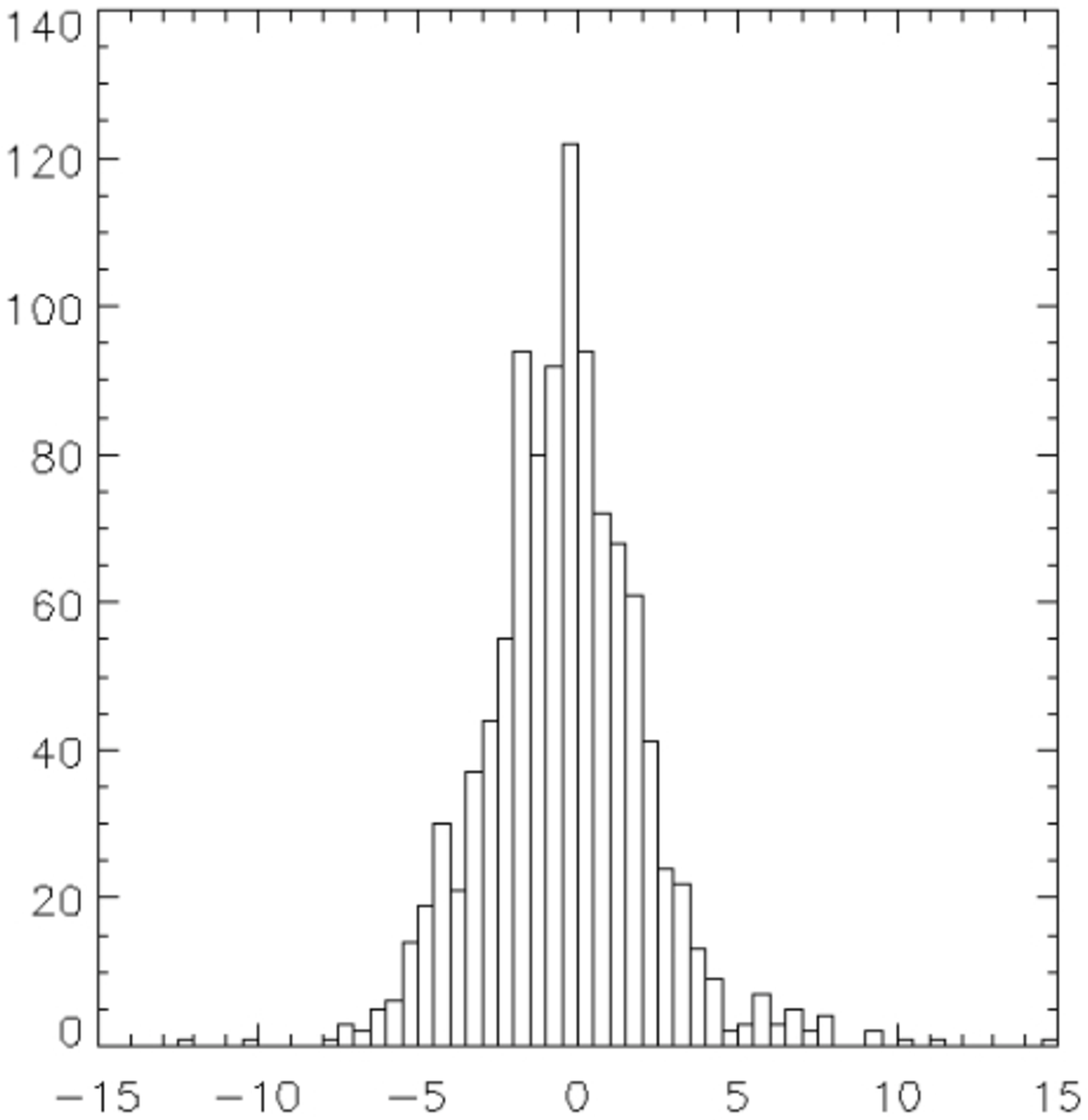}
\includegraphics[scale = 0.38]{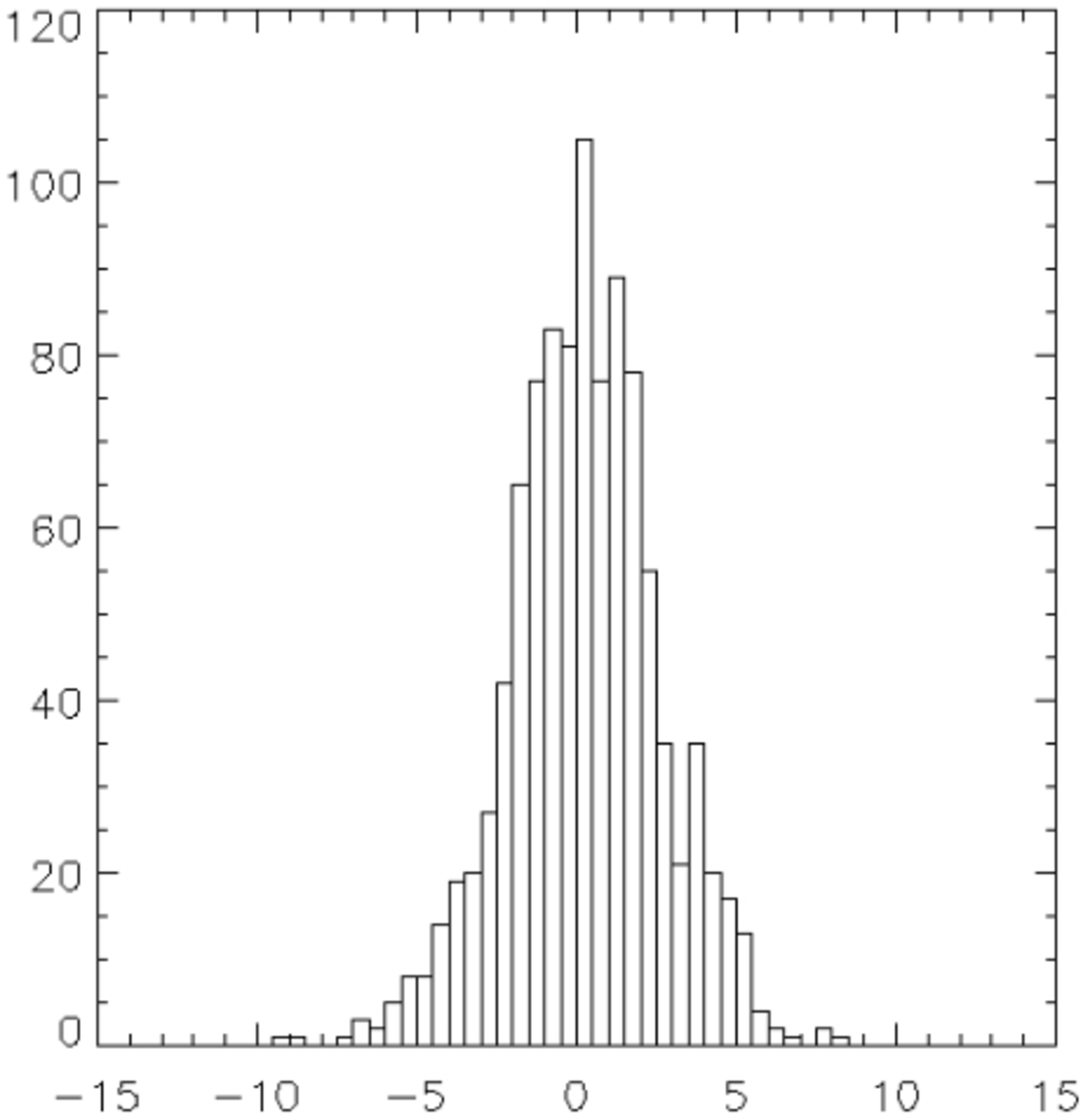}
\includegraphics[scale = 0.38]{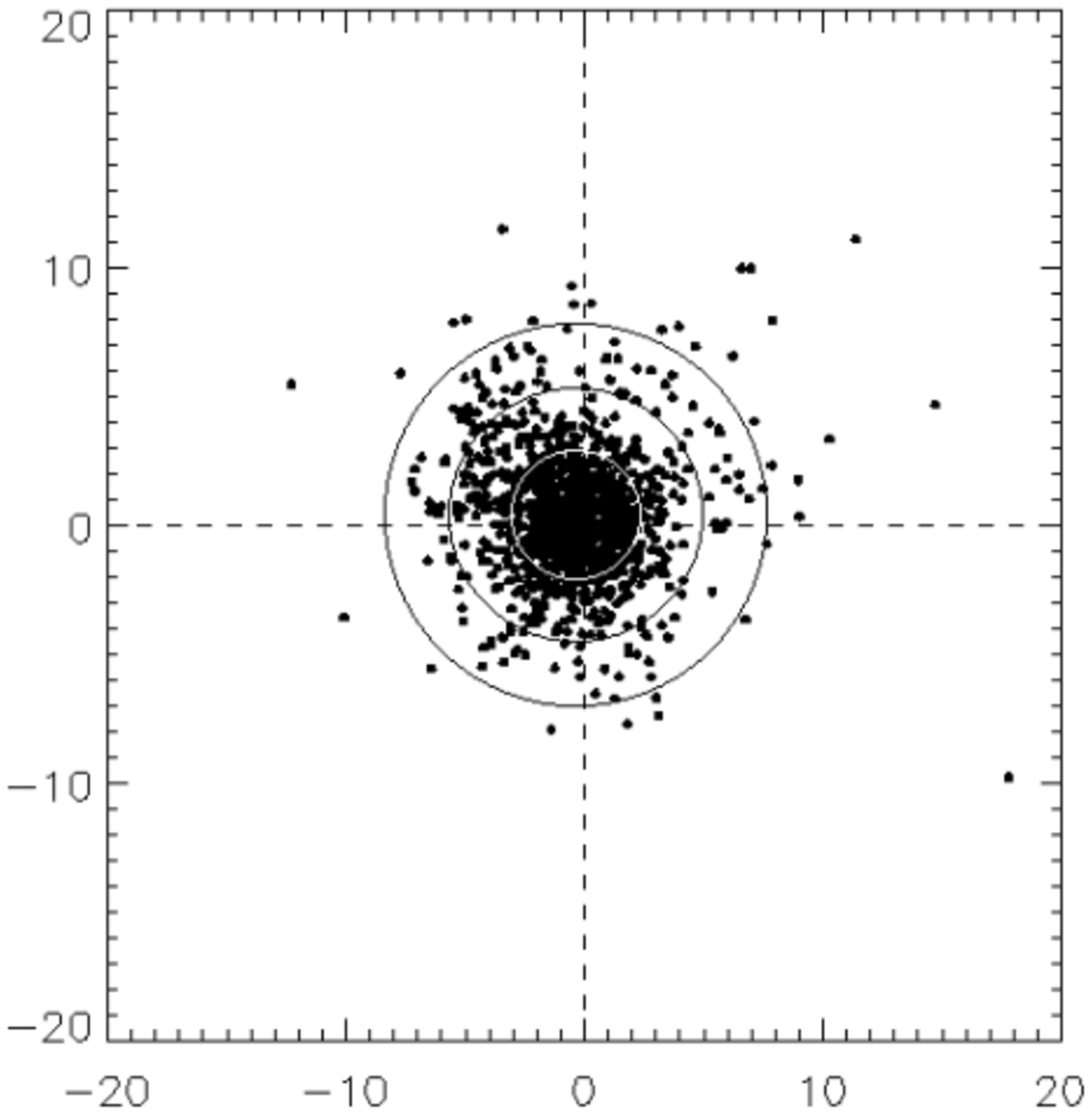}\\

\caption{Histogram of the differences in the mean proper motion from this work minus D14
  (first line), the values from this work minus M1 (second line), the values from this work minus M2 (third line), and the values from this work minus M3 (last line), in
  $\mu_{\alpha} \cos \delta$ (left panel) and $\mu_{\delta}$ (middle panel). 
In the right panel are given the differences and the circles show the
  regions of the one, two, and three standard deviations obtained in
  this study.}
  \label{fig:compares1}
\end{figure*}

\begin{figure*}
\centering
\includegraphics[scale = 0.38]{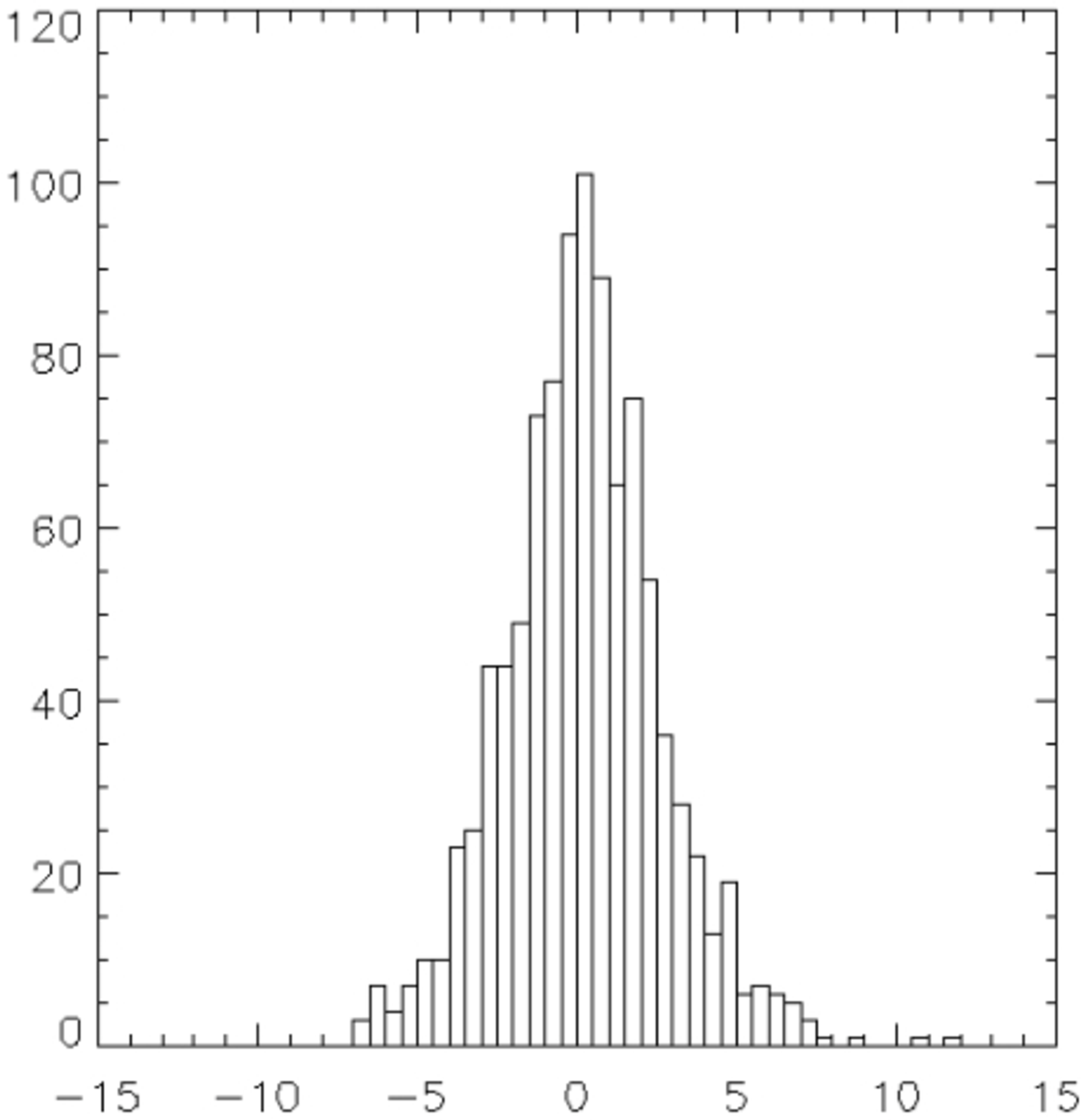}
\includegraphics[scale = 0.38]{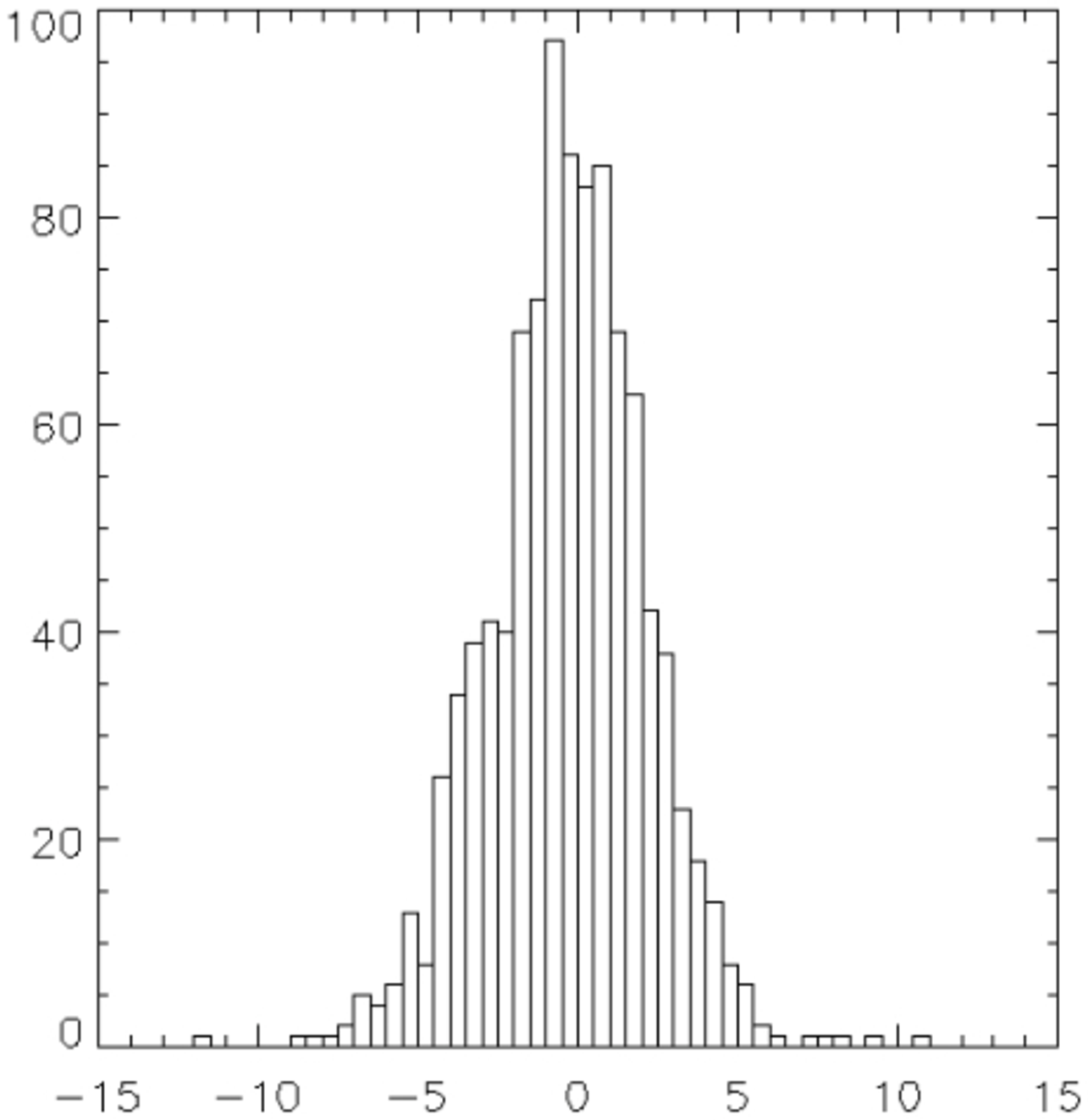}
\includegraphics[scale = 0.38]{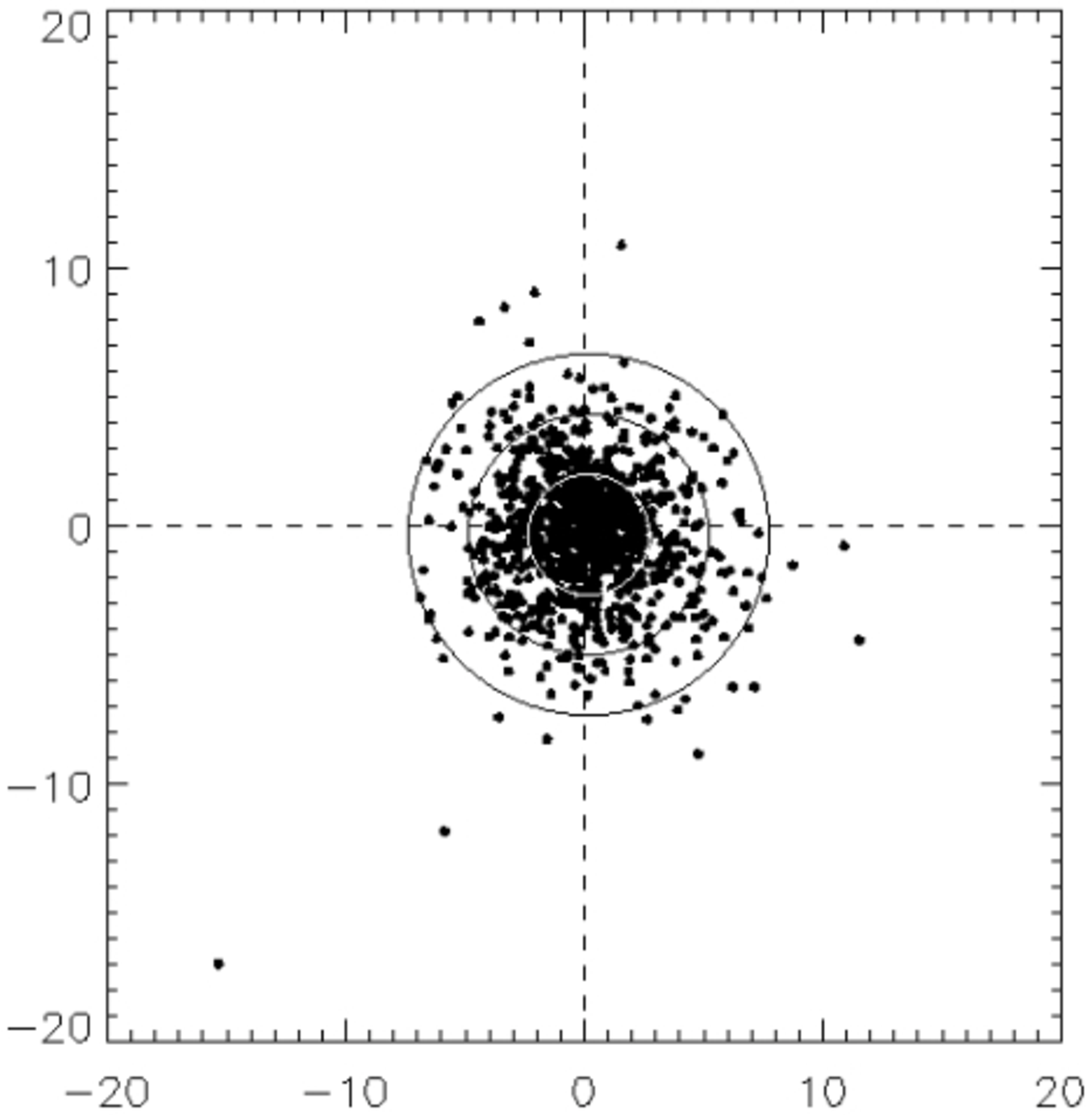}\\

\includegraphics[scale = 0.38]{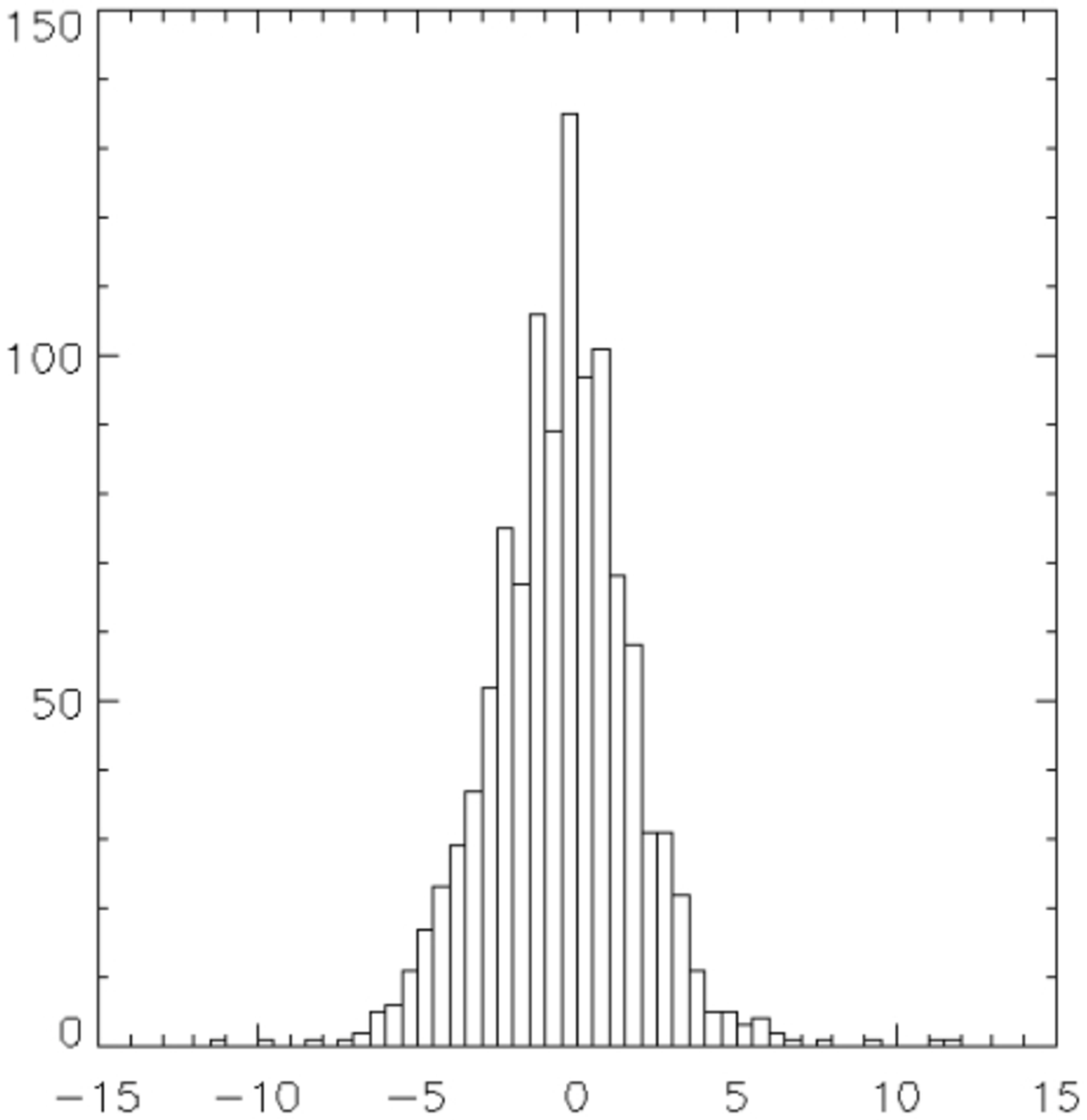}
\includegraphics[scale = 0.38]{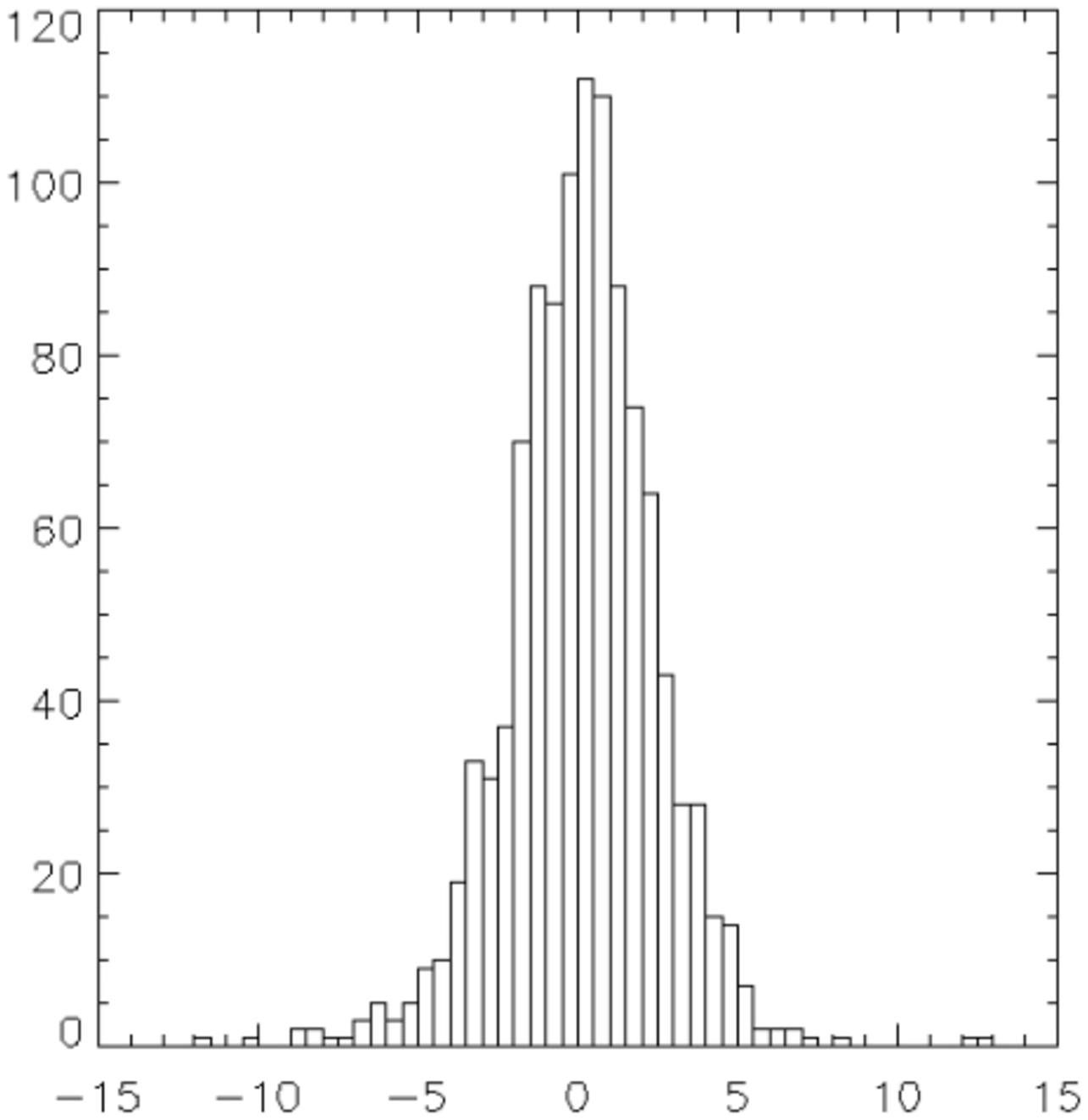}
\includegraphics[scale = 0.38]{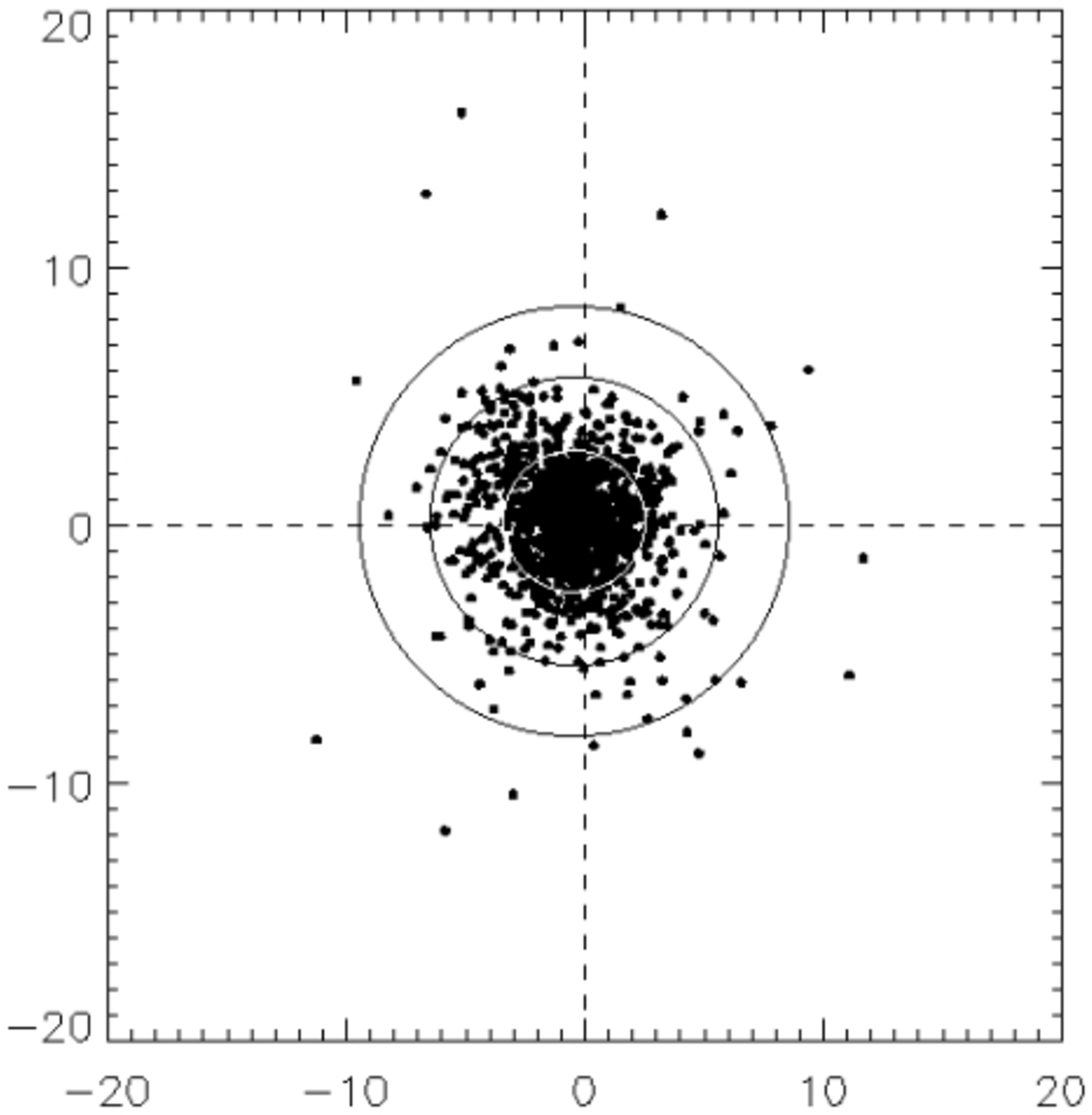} \\

\caption{Histogram of the differences in the mean proper motion from this work minus KA13
  (first line), and the values from this work minus DAML02 (second line) in
  $\mu_{\alpha} \cos \delta$ (left panel) and $\mu_{\delta}$ (meddle panel). 
In the right panel are given the differences and the circles show the
  regions of the one, two, and three standard deviations obtained in
  this study.}
  \label{fig:compares2}
\end{figure*}

Below we present a brief comment about the references used in the comparison with 
literature's results.

\subsection{Dias et al. 2014}
The comparison with D14 is interesting since we used exactly the same method 
and procedures applied to UCAC4 data. In total we found 1062 common objects.
In a statistical sense, the results obtained in this work were quite similar to those 
determined in D14.
The Gaussian fit to the differences (Fig. \ref{fig:compares1}) provided in the Table \ref{tab:compares}  
gives a mean proper motion difference 
of $-0.6 mas~yr^{-1}$ in $\mu_{\alpha} \cos \delta$ and $0.2 mas~yr^{-1}$ in $\mu_{\delta}$ ; 
the standard deviations are $2.0 mas~yr^{-1}$ in $\mu_{\alpha} \cos \delta$ and $2.2 mas~yr^{-1}$ in $\mu_{\delta}$. 
The differences between the mean proper motions obtained with UCAC5 and UCAC4  are very consistent. We note that the values
of the dispersions obtained in this work are in average $1.0 mas~yr^{-1}$ smaller than the values obtained with the UCAC4 data.

It is expected that the proper motions of the UCAC5 represent an improvement over the UCAC4 ones, particularly for magnitudes above R=14-15. However, we could not strictly verify this here. Our method weights the stars by their proper motions errors, and both in the UCAC4 and UCAC5 these errors steadily start to increase from magnitude R=14-15. Thus, fainter stars tend to be less weighted regardless of which UCAC catalog we use.

\subsection{Sampedro et al. 2017}

\citet{Sampedro2017}, investigated all clusters of DAML02 catalog and determined mean proper motion and membership using the UCAC4 data. The results were obtained by three different methods named M1, M2 and M3, briefly described below. For a complete description we refer the reader the original paper and references therein.

M1 is a method published by \citet{Sampedro2016}. 
It is a kind of geometrical method which estimates, in an N-dimensional space, 
the membership probabilities using the proper motion data by means of the distances 
between every star and the cluster central over-density, considering a mixture of two 
1-Dimensional Gaussians, one for the cluster and other for the field stars.

M2 is the nonparametric method published by \citet{Cabrera-Cano1990} which determines the members of the clusters using the positions (angular distances) and the proper motions data. It is interesting since it does not consider any a priori assumptions about the cluster and field star distributions. One advantage is that there is no assumed hypotheses for the variables used or their statistical properties since for many cases errors and bias are difficult to model by a parametric distribution.

M3 follows a parametric method published by \citet{Cabrera-Cano1985}. 
It is very similar to the classical Sanders method which is a overlapping normal 
bivariate frequency functions, one elliptical for the field and one circular for the cluster.

While our method uses the individual proper motions errors, M2 has the advantage of not using a Gaussian in the field stars model. We notice that this approach may be more appropriate since field star distribution is not Gaussian for many clusters.

The comparison of mean proper motions presented in Table \ref{tab:compares} and Fig. \ref{fig:compares1} shows that the results obtained by the different methods are very consistent. 

As shown by \citet{Balaguer2004} we notice that the parametric and non-parametric approaches give similar values of mean proper motions and generally provide same segregation of cluster and field populations.
The results published by \citet{Sampedro2017} confirm
it using a larger sample of clusters.

\subsection{Kharchenko et al. 2013}

The authors used the PPMXL catalog \citep{Roeser2010} and 2MASS \citep{Curtis2013} to determine kinematic and photometric membership probabilities for stars in the region of 3006 clusters.

Basically, the membership probabilities of stars were determined from their location with respect to the isochrones in photometric diagrams and the average cluster proper 
motion in kinematic diagrams.

\subsection{DAML02 Catalog}

The DAML02 catalog presents fundamental and kinematic parameters of 2167 know galactic open clusters. The last version 3.5 contains mean proper motions for $97\%$ of the sample. 
In short, the greatest contributions for mean proper motions in DAML02 catalog are the following: 1594 clusters from D14, 179 from the Khachenko's team \citep{Kharchenko2013,Kharchenko2012,Kharchenko2005}, 78 from \citet{Baumgardt2000} and 37 from \citet{Dias2001,Dias2002b}.
In the Table \ref{tab:compares} and Fig.\ref{fig:compares2} we show that the values of mean propers motion for 1102 clusters obtained in this study agree very well with the values given in DAML02 catalog. The dispersions of the differences are not very different of those obtained for homogeneous samples. 

For 112 common clusters, i.e. about 10\% of the cases, we found heterogeneous mean proper motion estimations with respect
to the last version of DAML02. The cases of this sample can be explained as follows. 
For 60 open clusters studied by \citet{Baumgardt2000} the 
mean proper motion were based on the Hipparcos proper motion and one to five stars.
For the other 52 objects which were investigated by different authors we verified that the mean proper motions were determined both using small number of stars and the older catalogs UCAC2 \citep{Zacharias2004} or UCAC3 \citep{Zacharias2010}, or the
TYCHO-2 \citep{Hog2000}. Here, for these 112 clusters we give mean proper motions in the GAIA reference frame based 
in a large number of members.

The open clusters Alessi 52, FSR 0647, FSR 0828, FSR 0814 and Dutra-Bica 50 had no proper motions listed in DAML02 catalog. For these five clusters this is the first 
determination of mean proper motion and membership probability. 

What happens to nearby clusters (32 clusters with distance less than 350 pc and diameter greater than 40 arcmin) is that the method cannot satisfactorily find the population of dozens of cluster stars in thousands stars in the sample.
It is possible only for cases like Pleiades, which stars have the proper motions very different from the field.

For those cases we note that our solutions agree relatively well with the values found in the literature, but on the other hand the number of stars considered as members is in excess with that expected by the mass function. This happens because the criterion of member selection is only the proper motion. All the stars with proper motions by chance "under" the Gaussian which defines the cluster population are considered as members. It is one of the main  drawbacks of the method which invariably implies in considering intruding field stars as cluster members.

In the sample of nearby clusters, the clusters Platais 5, Melotte 20, Platais 10, Melotte 25, Platais 2, Platais 9, Platais 8, and Collinder 285 have mean proper motions distinct from the field, but their diameters are greater than 300 arcmin. We chose not to analyze these cases for the reason explained above.
For the "small" ones, as the clusters Mamajek 2, Feigelson 1, Ruprecht 147, with large mean proper motion, the solution given by the method was inadequate because of the very small number of star members found, possibly by an excess of discarded outliers stars.

Finally, in our opinion, all these cases of nearby open clusters require membership determination using different datasets and not just proper motions. This is the great advantage of the GAIA data (besides unprecedented precision) over the limitations of the catalogs used so far.

\section{Conclusions}

Motivated by the improvement by a factor of 2 in the precision and accuracy of the proper motions with respect to the early UCAC4 catalog, and now in the GAIA reference frame, we used proper motion data from the new UCAC5 catalog to revisit our previous analyses of membership and mean proper motion of all cataloged open clusters.

Satisfactory results were obtained for a sample of 1108 open clusters, resulting in the most precise kinematic data set available up to date in the current GAIA reference frame.

For 112 open clusters we could update the mean proper motion in the DAML02 catalog using a large number of member stars with UCAC5 data.
For the clusters Alessi 52, FSR 0647, FSR 0828, FSR 0814 and Dutra-Bica 50 this is the first determination of a mean proper motion.

The results presented in this study indicates that the DAML02 catalog which presents a heterogeneous compilation of
mean proper motions are statistically similar to those obtained for homogeneous samples.

It is our last DAML02 update based on membership study using proper motion data only. In the forthcoming papers we will present memberships using kinematic, photometric and spectroscopic GAIA data.

\section*{Acknowledgements}
H. Monteiro would like to thank
FAPEMIG grants APQ-02030-10 and CEX-PPM-00235-12.
M. Assafin thanks the CNPq (Grants 473002/2013-2 and 308721/2011-0) and FAPERJ (Grant E-26/111.488/2013).
This research was performed using the facilities of the Laborat\'orio de Astrof\'isica Computacional da Universidade Federal de Itajub\'a (LAC-UNIFEI).
We employed catalogs from CDS/Simbad (Strasbourg).






\bibliographystyle{mnras}
\bibliography{refs} 




\bsp	
\label{lastpage}
\end{document}